\documentclass[trackchanges]{aastex701}
\usepackage{tikz} 
\usetikzlibrary{positioning}
\usepackage{amsmath}
\usepackage{graphicx}
\usepackage{siunitx,tabularx,ragged2e}
\usepackage{array}

\newcommand{\panelheight}{5.5cm}

\begin{document}

\title{Multimessenger constraints on the origin of long gamma-ray bursts with plateaus: three cases}

\author[orcid=0000-0003-1943-010X,gname='Aleksander',sname='Lenart']{Aleksander Łukasz Lenart}
\altaffiliation{The corresponding author}
\affiliation{Doctoral School of Exact and Natural Sciences, Jagiellonian University, Kraków, Poland}
\affiliation{Astronomical Observatory of Jagiellonian University in Kraków, Poland}
\email[show]{aleksander.lenart@doctoral.uj.edu.pl}  

\author[orcid=0000-0003-4442-8546,gname=Maria, sname='Dainotti']{Maria Giovanna Dainotti} 
\affiliation{Division of Science, National Astronomical Observatory of Japan, 2-21-1 Osawa, Mitaka, 181-8588 Tokyo, Japan}
\affiliation{The Graduate University for Advanced Studies (SOKENDAI), Shonankokusaimura, Hayama, Miura District, Kanagawa 240-0115, Japan}
\affiliation{Space Science Institute, 4765 Walnut St Ste B, Boulder, CO 80301, USA}
\email{maria.dainotti@nao.ac.jp}

\author[orcid=0000-0002-9212-411X]{Maurice van Putten}
\affiliation{Department of Physics and Astronomy, Sejong University, 98 Gunja-Dong, Gwangjin-gu, Seoul 143-747, Republic of Korea}
\email{}

\author[orcid=0000-0002-7025-284X]{Shigehiro Nagataki}
\affiliation{RIKEN Center for Interdisciplinary Theoretical and Mathematical Sciences (iTHEMS), Wakō, Saitama, 351-0198, Japan}
\affiliation{Astrophysical Big Bang Laboratory (ABBL), RIKEN Pioneering Research Institute (PRI), Wakō, Saitama, 351-0198, Japan}
\affiliation{Astrophysical Big Bang Group (ABBG), Okinawa Institute of Science and Technology Graduate University (OIST), 1919-1 Tancha, Onna-son, Kunigami-gun, Okinawa904-0495, Japan}
\affiliation{RIKEN-Berkeley Center, RIKEN iTHEMS, University of California, Berkeley, Berkeley, CA94720, USA}
\email{}

\author[orcid=0000-0002-0173-6453]{Nissim Fraija}
\affiliation{Instituto de Astronomía, Universidad Nacional Autónoma de México, Circuito Exterior, C.U., A. Postal 70-264, 04510, México City, Mexico}
\email{}

\author[orcid=0000-0003-3358-0665,gname=Agnieszka,sname=Pollo]{Agnieszka Pollo}
\affiliation{Astronomical Observatory of Jagiellonian University in Kraków, Poland}
\affiliation{National Centre for Nuclear Research, Świerk, Poland}
\email{agnieszka.pollo@uj.edu.pl}

\author[orcid=0000-0002-8028-0991]{Dieter H. Hartmann}
\affiliation{Department of Physics and Astronomy, Clemson University, Clemson, SC 29634, USA}
\email{}

\begin{abstract}
The central engines of long gamma-ray bursts (LGRBs) with X-ray plateaus remain uncertain. They involve a magnetar or black hole (BH), powered by spin energy and/or accretion.
We examine how gravitational waves (GWs) discriminate between these scenarios.
For collapsars, we consider three models: (A) a prompt phase powered by hyper-accretion onto a BH, followed by a spin-down-powered plateau; (B) a prompt phase powered by BH spin-down; and (C) a magnetar powering both the prompt and plateau phases.
Each predicts a distinct GW signature in timing, energy, and frequency, ranging from broadband emission 
to a coherent descending chirp.
We fit the selected X-ray plateaus of Swift GRBs using the model (A).
For GRB~080319C, GRB~111209A, and GRB~130925A, we obtain BH masses of $M_{\rm BH}\sim24$--$124\,M_\odot$, with disc masses of $\sim10\%\,M_{\rm BH}$ at the end of the prompt phase.
Mass-scaling of numerical simulations of disc emission yields GW strains of $\sim1.5-1.8\times10^{-24}$ at few\,hHz.
This emission should be detectable by Cosmic Explorer and the Einstein Telescope, with an inferred rate from $0.5$ to $1.5$\,yr$^{-1}$.
However, ultra-long GRBs powered by tidal disruption events are expected to emit GWs at frequencies ranging from millihertz to decihertz, potentially within the sensitivity range of DECIGO and/or Big Bang Observer (BBO).
Current or future detectors may detect GWs associated with BH-powered GRB, but magnetar engines, require next-generation detectors.
Current and future detectors could map LGRB central-engine activity.
\end{abstract}



\section{Introduction} 

Gamma-ray bursts (GRBs) are among the most energetic 
and relativistic transients, releasing  
extreme amounts of energy within timescales ranging from milliseconds to several minutes \citep{Kumar2015,Piran2004}. 
Current models suggest that GRBs are powered by compact objects -- neutron stars (NSs) or black holes (BHs) -- formed 
in catastrophic events such as the merger of
an NS with a companion NS or BH or the core-collapse of massive stars \citep{Woosley1993,Eichler1989}. 
These processes are thought to produce either a rapidly rotating BH surrounded by an accretion disk and/or torus, or a highly magnetized 
and possibly rapidly spinning NS (magnetar).
Both are believed to launch ultra-relativistic baryon-poor jets responsible for the observed gamma-ray emission 
\citep[e.g.][]{Metzger2011, Lowell_2024}. 
Many models have been
proposed suggesting GRBs as potentially powerful sources of detectable gravitational waves (GWs),
which promises to unveil the physical nature of their central engines \citep{cut02}.
This multi-messenger outlook manifested in the double NS merger event GW170817 
with compelling observational evidence for a newborn BH GRB central engine based on the associated kilonova and GW data
\citep{Abbott2017GW170817,Abbott2017MultiMessenger,pod04,van19a,ram19,van23}) -- presently, the only confidently detected source of this kind.

GRBs produced by the core-collapse of massive stars (collapsars) are also believed to be potentially promising sources of GWs. 
Here also, a GW detection or confident null detection may unveil the nature of the central engine -- a NS or BH \citep{vanPutten2024} possibly including the associated state of accretion.
For the case of a NS, \cite{CorsiMeszaros2009} discuss possible GW emission by the bar-mode instability of a newly formed NS. 
For the case of a BH. \cite{Gottlieb2024ApJ...972L...4G} recently simulated the broadband GW emission
generated by the Rossby instability in an accretion disc during a state of hyper-accretion. \cite{vanPutten2001b,vanPutten2003,vanPutten2004} describe 
the potential of descending chirps--coherent GWs with frequency decreasing over time from an inner disk or torus in catalysis of BH-spin energy during the prompt GRB phase.

The frequencies, energetics, and spectral morphology of GW emission in these scenarios differ appreciably, significant to the prospect of unveiling the nature of GRB central engines
\citep{cut02}, which possibly extends to the pre- and post-GRB phase currently probed solely in the electromagnetic spectrum.
For the collapsar scenario considered here, a detection of the GW emission
could finally identify
this central engine to be a BH or a magnetar.
This question, potentially extending to the question on the physical trigger of the explosion and post-GRB emission seen in X-ray plateaus, 
is a holy grail of the GRB field, 
which remains unanswered for almost 30 years since the first core-collapse event SN 1998bw firmly associated with GRB \cite{Galama1998Natur.395..670G} and the first paper proposing a magnetar central engine for GRBs \citep{DaiLu1998MNRAS}.

For collapsars, we also note
more exotic models of GW production
involving the formation of subsolar NSs \citep{ims88,Chen_2025}
of mass ($\sim0.1\rm\,M_{\odot}$) in fragmented accretion discs 
around a more massive central BH or NS. 
If subsequent inverse mass transfer does not lead to an explosion, this low-mass NS may merge with the central object, revealed by a characteristic merger chirp in GWs.
\cite{Lev15,ShibataPhysRevD.103.063037} discuss a possible formation of a spiral-arm structure in the accretion disc, efficiently producing GWs.
If the central object is a newborn magnetar,
it may experience a transition from a Jacobi ellipsoid to the Maclaurin spheroid \citep{Rueda2022PhRvD.106h3004R}. 
These diverse scenarios highlight the need for model-independent probes of core-collapse events for their multi-messenger output. Here, our primary focus is on possible signatures in broadband disk emission, accompanying LGRBs with X-ray plateaus.

In LGRBs with X-ray plateaus, the activity of the central engine may extend from a pre-GRB phase, through the prompt GRB phase and well into the post-GRB phase including the powering of X-ray plateaus. The question on the central engine, therefore, is broader than the prompt GRB phase, and extends to   
GRB afterglow studies and, ultimately, to its final remnant.
Frequently observed are GRB afterglows featuring energetic and long-lasting X-ray luminous plateaus of duration $\sim10^2$--$10^5$ s 
\citep{Nousek2006,Zhang2006}, first discovered by \textit{Swift} \citep{Gehrels2004}.
Interestingly, both the plateau features and the Dainotti relation \citep{Dainotti2008,dainotti2010,Dainotti2013,dainotti2011a}, between the X-ray plateaus and their rest-frame duration \citep{Dainotti2008,Dainotti2011} can be interpreted within two main scenarios: a sustained activity of the central engine, 
commonly attributed to continuous energy injection from a rapidly rotating magnetar \citep{DaiLu1998MNRAS,zhangmeszaros2001,Metzger2011,Rowlinson2013,Rowlinson2014,Rea2015,Gompertz2015MNRAS.448..629G,stratta2018} or to a BH \citep{Cannizzo2009,Cannizzo2011,Lenart2025}. 
By energy, many plateau observations can be explained within the magnetar framework.
Some exceed the maximal spin-energy of a canonical millisecond magnetar, 
suggesting the potentially more powerful involvement of a BH \citep{Li2018b,Lu2015}.
These open issues extend to more recently observed peculiar GRBs, i.e.
ultra-long GRBs (ULGRBs)
with durations (observer frame period over which 90\% of energy is detected) $T_{90}>10^3$s \citep{Gendre2013,Levan2014,Gendre2019}.
Their astronomical origin may be found in, e.g., core-collapse of very large stars \citep{Levan2014}, blue supergiants \citep{Gendre2013,Nakauchi2013ApJ...778...67N,Ioka2016ApJ...833..110I} or Population III stars \citep{Liu2018ApJ...852...20L}, alternatively, in tidal disruption events, TDEs \citep{Levan2014,Ioka2016ApJ...833..110I}.

\cite{dainotti2017,Dainotti2020ApJ...904...97D} found that the scatter of ULGRBs relative to the Dainotti correlation of normal LGRBs is broader than in the regular LGRB sample \citep{Dainotti2008,dainotti2010,Dainotti2013}, suggesting multiple progenitor channels.

The complex sequence of energetic phases before, during and after a GRB poses novel observational constraints on the central engine, on the nature of the compact object -- NS or BH --- and its state of accretion, which may include

\begin{itemize}
\item 
    Hyper-accretion, powering the prompt phase that tends to spin up the BH, possibly followed by a plateau phase, sustained by magnetic winds sourced by a central energy reservoir in angular momentum \citep{bis70}, accelerating accretion and/or spinning down the compact object \citep{Lenart2025};
    \item Spin-down during the prompt GRB phase, while the plateau phase is powered by its compact remnant through residual accretion and/or its spin, powering a relatively less luminous jet \cite{}.
    \item An NS central engine during the prompt GRB phase, followed by gravitational collapse to a BH central engine powering the plateaus phase \citep{Rowlinson2014, Rea2015, stratta2018}.
\end{itemize}

Based on some specific models for  multi-messenger emission specific to events featuring X-ray plateaus, we here unveil these different scenarios by GW-event timing \citep[e.g.][]{van23}.
Of particular interest are constraints on BH masses $M$ inferred from these plateaus. \cite{Lenart2025} constructed a model of BH spin-down-powered afterglow, predicting $M\sim 1-30\,\rm M_{\odot}$ with a peak around $M\simeq 7\,\rm M_{\odot}$, while a few of the most extreme events point to masses up to $M\sim 100\,\rm M_{\odot}$. 
This analysis gives a consistent reconstruction of the Dainotti correlation \citep{Dainotti2008,Dainotti2011} and its scatter \citep{Lenart2025}. 
 
Presciently envisioned earlier \citep{cut02}, the present transition of LIGO-Virgo from detectors to observatories opens a radically new avenue to unveil the physical triggers of GRBs.

The multi-messenger prospects of LGRBs with plateaus outlined above offer a unique opportunity to probe these relativistic events at their core, 
from formation and evolution of their central engines through a complex evolutionary process, from gravitational core-collapse of massive stars to their final remnants.
In \S\ref{sec:data}, we review the leading models of LGRBs and discuss how current and future GW observations can identify different scenarios, promising new insights into the physics and origins of compact objects. 
In \S\ref{sec:model}, we demonstrate how multi-messenger analysis of electromagnetic and gravitational radiation can constrain key physical parameters. 
We analyse past GRBs and assess the possible future observability of their associated GWs in \S\ref{sec:detectability}. We discuss the GW constraints on GRB physics in \S\ref{sec:constraints}.

\section{Data sample} \label{sec:data}

Since our light curve (LC) model \citep{Lenart2025} requires the presence of the plateau, we gathered the 256 GRBs with well-defined plateaus (well fitted by the \cite{Willingale2007} model; for details, see \cite{Narendra2024}) observed by the Swift Neil Gehrels Observatory between 2005 and 2023. 

We further restrict our sample to the four GRBs with the highest plateau energy-to-proper-distance ratio ($>3\times 10^{23}\,\rm erg\, cm^{-1}$) (as they have the highest chance of producing detectable GWs; see Sec. \ref{sec:detectability}). We provide their IDs, redshifts ($z$), prompt isotropic energy ($E_{iso}$), prompt spectrum peak energy ($E_{peak}$), and information on the SNe association and optical plateau detection in Tab. \ref{tab:data}.

\begin{table*}[ht!]
    \centering
    \begin{tabular}{|c|c|c|c|c|c|c|}
    \hline
        GRB & $z$ & $T_{90}\,\rm [s]$\footnote{Taken from \textit{Swift} BAT UKSSDC database} & $E_{iso}\,\rm [erg]$ & $E_{peak}\,\rm [keV]$ & Associated SNe & Optical plateau detection \\\hline
        080319C & 1.95 & $33.9\pm 8.9$ & $1.05^{+0.24}_{-0.15}\times 10^{53}$ \footnote{\label{f:1}\cite{Golenetskii2008GCN..7487....1G}} & $307^{+141}_{-92}$ \footref{f:1} & No data & Yes \\\hline
        111209A & 0.677 & Unconstrained & $(5.82\pm 0.73)\times 10^{53}$ \footnote{\label{f:2} \cite{Golenetskii2011GCN.12663....1G}} & $520\pm89$ \footref{f:2} & Super-luminous \footnote{\cite{Kann2019}} & Yes \\\hline
        130925A & 0.347 & $161.1\pm 3.5$ & $(2.9\pm0.3)\times 10^{53}$ \footnote{\label{f:3}\cite{Evans2014}} & $65^{+13}_{-16}$ \footref{f:3} & Upper limit\footnote{\cite{Greiner2014}} & Yes \\\hline
        211024B & 1.1137 & $603\pm 21$ & Unconstrained & Unconstrained & No data & Yes \\\hline
    \end{tabular}
    
    \caption{Properties of GRBs in our sample. The optical plateau data can be found in \cite{Dainotti2024opt}.}
    \label{tab:data}
\end{table*}

\section{Collapsar BH models} \label{sec:model}

\begin{figure*}[t] 
\centering 
\begin{tikzpicture}[ image/.style={ inner sep=0pt, anchor=north west }, textpanel/.style={ draw, align=left, text width=0.5\textwidth, minimum height=\panelheight, inner sep=8pt, anchor=north west } ] 
\node[image] (img1) at (0,0) {\includegraphics[width=0.5\textwidth,height=\panelheight]{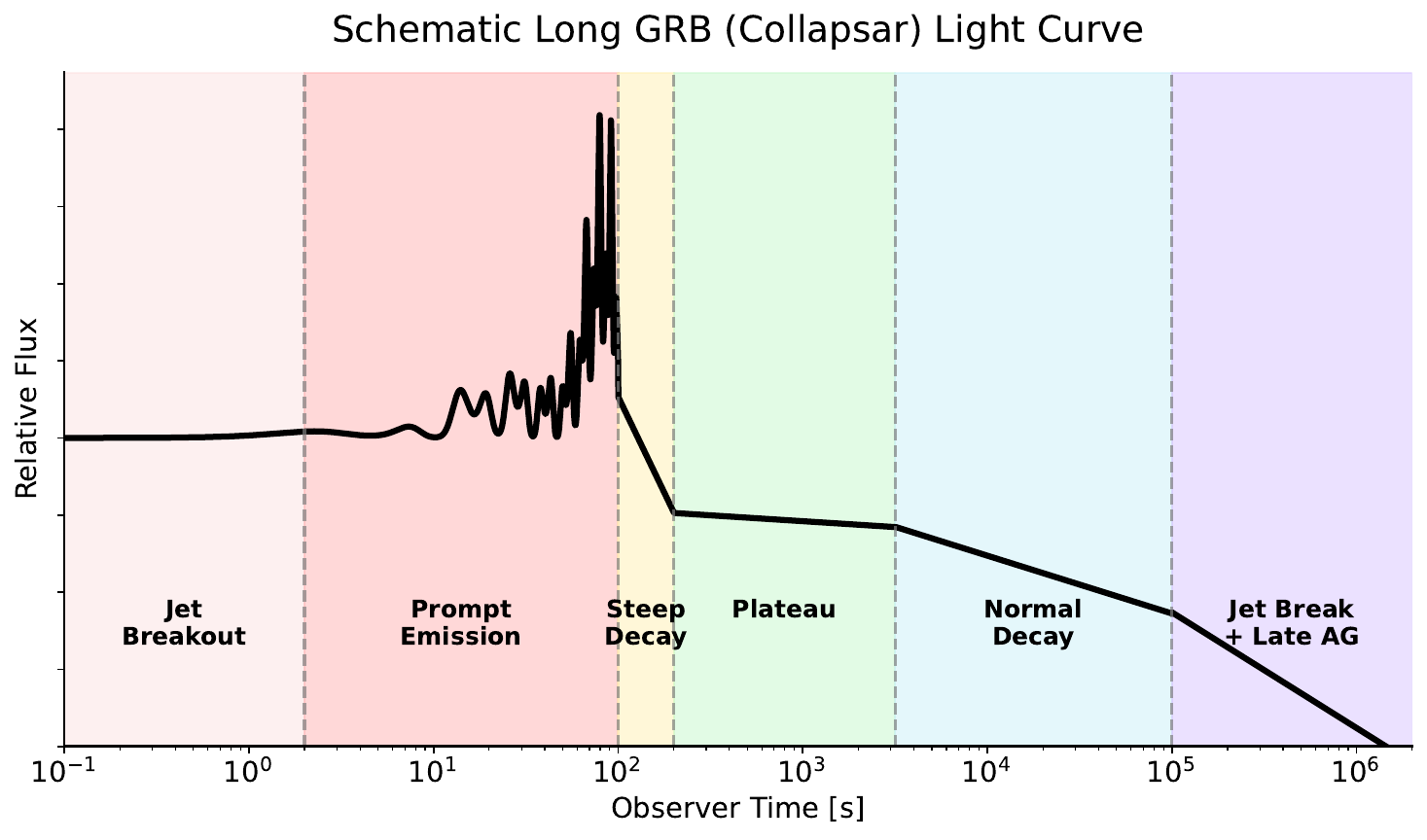}}; 
\node[textpanel,right=0.cm of img1.north east,anchor=north west] (txt1) { 
\textbf{The standard GRB LC}\\ {\bf Jet breakout} marks the jet's emergence from the progenitor star, followed by the highly variable {\bf prompt emission}. The subsequent {\bf steep decay} is typically attributed to high-latitude emission \citep{Willingale2010MNRAS.403.1296W}, or some transition in emission mechanism, while the {\bf plateau} may indicate continued energy injection into the forward shock. The {\bf normal decay} reflects the standard afterglow evolution \citep{Zhang2006}, followed by a possible {\bf jet break and late afterglow}, when relativistic beaming effects diminish, producing a steeper decline that traces the deceleration of the blast wave.}; 
\node[image,below=0.3cm of img1.south west,anchor=north west] (img2) {
\includegraphics[width=0.5\textwidth,height=\panelheight]{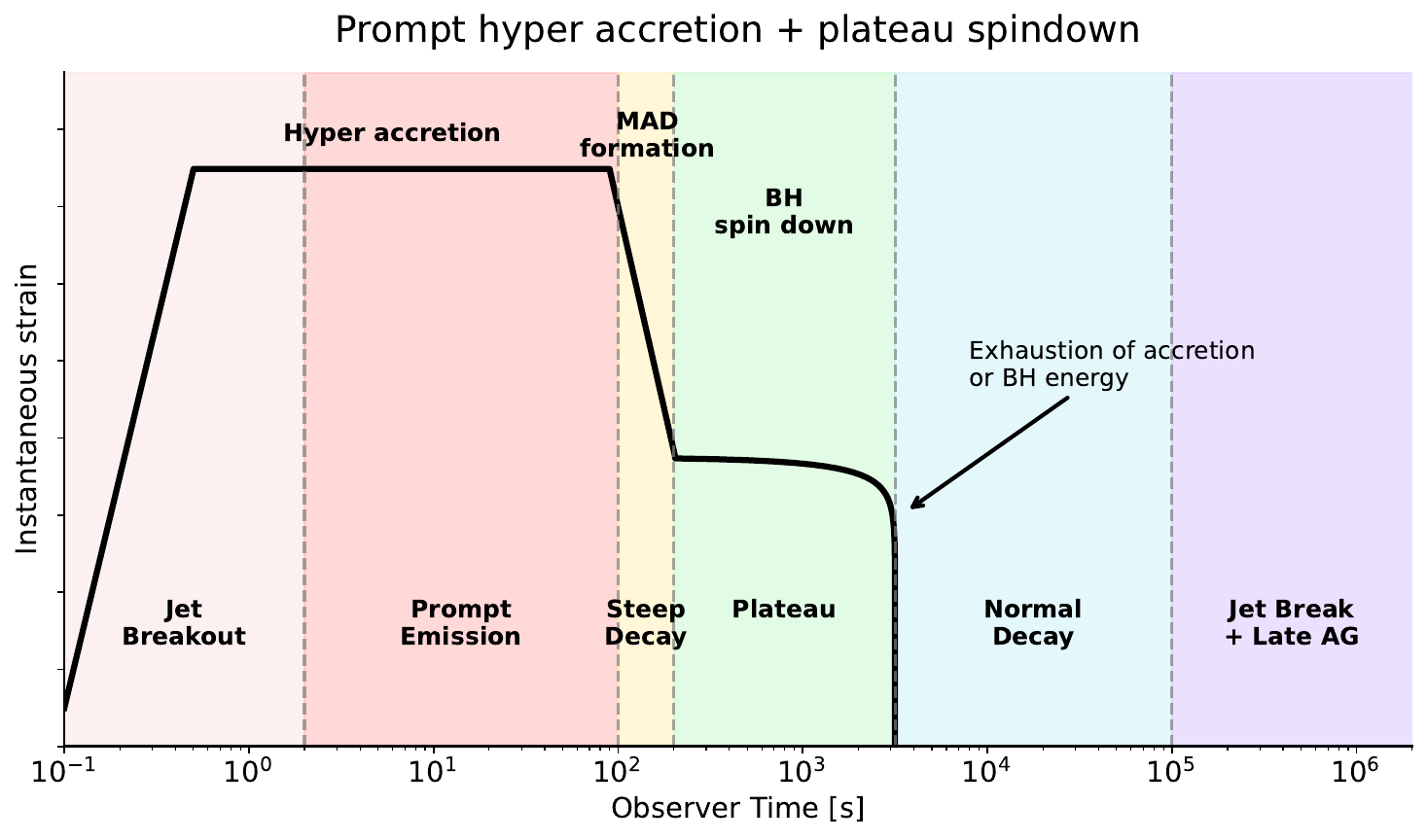}}; 
\node[textpanel,right=0.cm of img2.north east,anchor=north west] (txt2) { 
\textbf{Model A: Spin-down-powered plateau}\\ \cite{Gottlieb2024ApJ...972L...4G} analysed magneto-hydrodynamic simulation, where the dominant GW source is the turbulent disc formed right after the core collapse. The asymmetric mass motions in the disc generate broadband GWs in the LIGO band ($\sim 0.1{-}0.6$ kHz) over the timescale of $\sim10{-}100$ s, corresponding to the prompt GRB phase. The strain is a stochastic-like burst whose amplitude tracks the evolution of the disc. When the magnetic field increases and the MAD stage forms, the viscosity in the disc grows, spreading it and smoothing it; consequently, at this stage the BH can lose spin \citep{Lowell_2024} and power the plateau \citep{Lenart2025}. The BH can further produce GWs as long as it has an accretion disc; however, with much smaller efficiency.}; 
\node[image,below=0.3cm of img2.south west,anchor=north west] (img3) {
\includegraphics[width=0.5\textwidth,height=\panelheight]{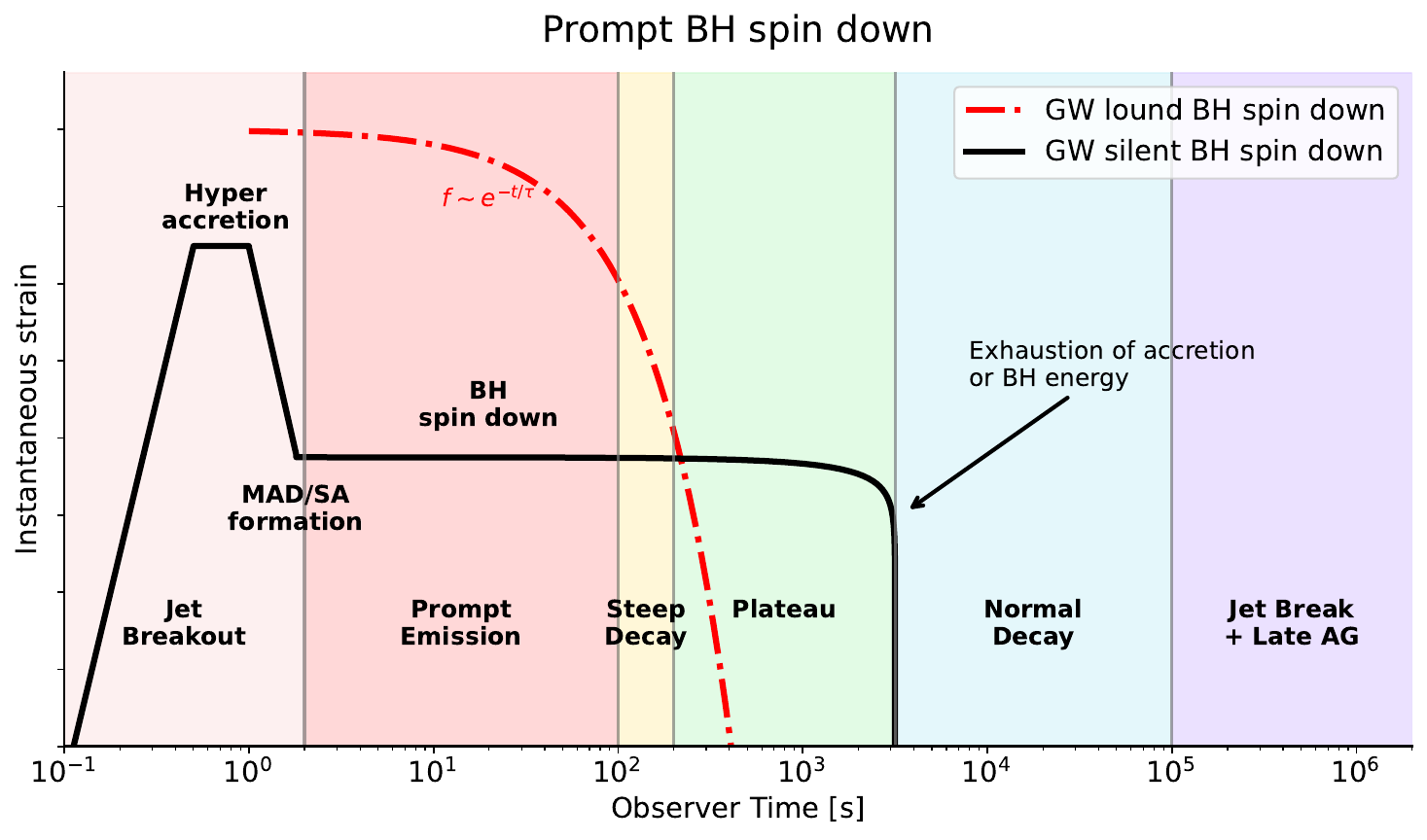}}; 
\node[textpanel,right=0.cm of img3.north east,anchor=north west] (txt3) { 
\textbf{Model B: Spin-down-powered prompt}\\ If the magnetic field can accumulate rapidly around the BH. Accretion flow may be arrested, by receiving energy-angular momentum from the BH (SA) over an inner-torus magnetosphere \citep{vanPutten2024} or by magnetic pressure (MAD).
In the case of SA, the BH can emit coherent GWs very efficiently, producing the descending chirp (GW-loud BH spin-down).
As the BH spin-energy reservoir is exhausted and the SA state ends, the GW strain fades away, terminating on the same timescale as the GRB prompt. In the MAD scenario, the disc is likely to be less turbulent and should produce low-luminosity broadband GW emission (GW-silent BH spin-down). The emission would continue past the spin-down stage as long as the BH has an accretion disc.
}; 
\node[image,below=0.3cm of img3.south west,anchor=north west] (img4) {
\includegraphics[width=0.5\textwidth,height=\panelheight]{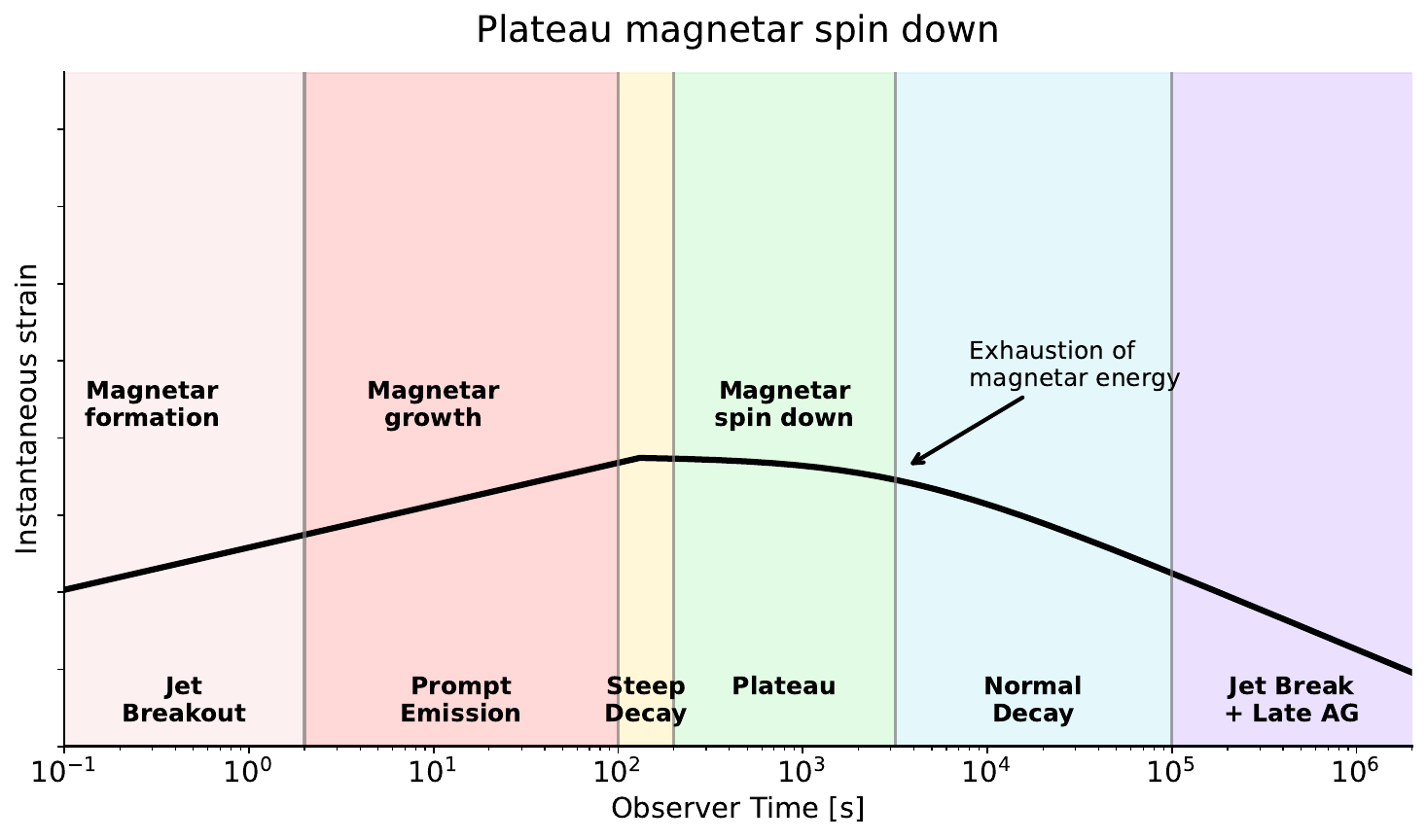}}; 
\node[textpanel,right=0.cm of img4.north east,anchor=north west] (txt4) { 
\textbf{Model C: Magnetar spin down}\\
In the \cite{CorsiMeszaros2009} model, a rapidly rotating magnetar initially produces only weak GW emission because the secular bar-mode instability has not yet fully developed.
During the prompt GRB phase, the deformation grows due to accretion on a newborn NS. The GW strain amplitude h(t) increases and reaches its maximum shortly after the GRB prompt.
During the plateau phase, the magnetar loses rotational energy through both electromagnetic and GW torques, so the GW frequency, $f_{\rm GW}\simeq 2f_{\rm spin}$, slowly decreases with time, producing a descending chirp.
The strain amplitude peaks early in the plateau and then gradually decreases as the star spins down, and the bar-mode weakens.
The overall waveform is a long-duration transient lasting minutes to about an hour.}; 
\end{tikzpicture} 
\caption{The first panel presents a typical GRB LC with the distinct phases marked with dashed grey lines and colors. The same phases are shown in all the other panels as a point of reference on the GW-GRB relation. The following panels show the cartoon visualization of expected characteristic strain behaviour over time. All axes are in log scale.} 
\label{fig:overview} 
\end{figure*}

\subsection{BH evolution}
\label{sec:GRBLC}

The energy and angular momentum conservation equations for BHs govern their time evolution. The BH mass ($M_{BH}$) and its spin ($a$) can be increased by the infalling matter from an accretion disc; on the other hand, a high magnetic field can exert a torque ($T_B$) on the spinning BH and extract its energy via open outflows in processes like \cite{BlandfordZnajek}. 
One may consider the set of differential equations:
\begin{equation}
\begin{array}{l}
    \frac{d M_{BH}}{dt} = e_{in}\dot{M} - \frac{L_{BZ}}{c^2}\\
    \frac{da}{dt} = \frac{c}{GM_{BH}^2}(l_{in}\dot{M} - T_B) - \frac{2a}{M_{BH}}\left(e_{in}\dot{M} - \frac{L_{BZ}}{c^2}\right),
\end{array}
\label{eq:BHevol}
\end{equation}
where $\dot{M}$ is the accretion rate, $L_{BZ}$ is the \cite{BlandfordZnajek} luminosity, and $e_{in}$ and $l_{in}$ are the efficiencies of energy and angular momentum transfer, respectively; $c$ is the speed of light and $G$ is the gravitational constant.
Several accretion regimes have been proposed for GRBs:

\begin{itemize}
    \item {\bf Hyper-accretion}: a very efficient accretion with high values of $e_{in}$ and $l_{in}$, which should follow relations derived by \cite{Bardeen1970}.
    \item {\bf Magnetically Arrested Disc (MAD) accretion}: the pressure from the infalling matter is in equilibrium with the magnetic field pressure. The magnetic field causes turbulence in the accretion disc and generates relatively small values of $e_{in}$ and $l_{in}$ \citep{Lowell_2024,Narayan2022}.
    \item {\bf Suspended Accretion (SA)}: 
    Surrounding high-density matter catalytically converts BH spin-energy into multi-messenger emission, mostly as gravitational radiation \citep{vanPutten2003,vanPutten2024}.
\end{itemize}

Given that the high magnetic field cannot be created immediately, there should be a period of very efficient accretion, i.e., hyper-accretion. Therefore:

\begin{enumerate}
    \item the BH should inherit a significant portion of angular momentum from its progenitor star, and
    \item hyper-accretion should further efficiently deliver angular momentum.
\end{enumerate}

Thus, the GRB-origin BH should be high-spinning right after its formation \citep{Li2024,Lei_2017}.
This period also matches very well the GRB phenomenology; the initial GRB prompt emission is very luminous and chaotic; thus, it is likely driven by the period of very efficient accretion. Alternatively, it should be powered by a very efficient BH energy extraction process like the one presented in \cite{vanPutten2012} in the case of SA or \cite{Issa2025} in the case of MAD.

At a given amount of time after the core collapse, it is expected that a very high magnetic field accumulates around the BH. This is attributed partially to the conservation of the magnetic field flux; when the star collapses to a very small size, the magnetic field is compressed and can become orders of magnitude larger than the initial stellar field. Therefore, it is likely that at a given stage of GRB's evolution, a MAD or SA will form. 

If the whole prompt phase is powered by hyper-accretion, when the MAD stage of the accretion is finally achieved, it is possible for the luminosity to drop significantly, thus ending the prompt phase \citep{Lenart2025}.
In the MAD scenario, the $l_{in}$ is relatively small \citep{Lowell_2024}, and in Eq. \ref{eq:BHevol} the magnetic torque and energy extraction factors can dominate, leading to a spin-down of a BH to a value as small as $a_{final} = 0.07$ \citep{Lowell_2024} or even $a_{final} < 0.035$ \citep{Narayan2022}. 
One can solve the set of Eq. \ref{eq:BHevol} for such a mode of accretion and compute the jet luminosity $L_{BZ}$ as a function of time, obtaining a broken power-law-like shape \citep{Lenart2025}. 
This theoretical light curve LC matches very well the shape of GRB LCs, as we show further in the appendix in Fig. \ref{fig:grblc}. Therefore, fitting this model to real data enables an estimation of the mass of the BH right at the beginning of the plateau emission, created in a GRB, and the accretion rate (See Sec. \ref{sec:electromagnetic}). We refer later to this scenario in relation to GWs, and call it {\bf Model A}.

Alternatively, if the prompt is powered by either MAD or SA states, the following plateau emission can be produced by the residual emission from the accretion disc, as discussed in \cite{Kumar2008,vanPutten2009arXiv0905.3367V}, if the whole disc was not swallowed by the BH. Although it is unknown how realistic this scenario is, given that the emission from the late disc would need to be collimated and produce a spectrum compatible with the synchrotron emission. Yet another scenario is that the BH would not lose a large portion of the spin during the prompt, but it would stabilise at a value, which would still allow production of a less luminous jet. Those scenarios have in common that the hyperaccretion should happen before the prompt emission, and the BH spin-down would power the prompt. We refer later to those scenarios as {\bf Model B}.

\subsection{Gravitational Waves}

\subsubsection{Broadband disc emission}

\cite{Gottlieb2024ApJ...972L...4G} demonstrated with simulations that the system of a newborn BH surrounded by an accretion disc can emit broadband GWs, thanks to the Rossby instabilities in the disc. The approximate formula for the characteristic strain at 200 Hz ($h_{200Hz}$, frequency in the source rest frame) takes the form:

\begin{equation}
    h_{200Hz} \approx 7\times 10^{-23} \epsilon(1+z)\frac{M_{BH}}{10\rm\, M_{\odot}}\frac{M_{disc}}{0.1\rm\, M_{\odot}}\frac{10\rm\, Mpc}{d_L}\frac{100\rm\, km}{R_{disc}},
    \label{eq:GWs}
\end{equation}

where $M_{disc}$ is the mass of the accretion disc, $R_{disc}$ is its radius, $d_L$ is the luminosity distance, and $\epsilon$ is the efficiency factor, which can take values between 0.1 and 0.2, depending on whether the accretion disc is weakly or strongly cooling, respectively, or even 0.3 in a very efficient scenario \citep{Gottlieb2024ApJ...972L...4G}. The authors provided the formula for characteristic strain at 200 Hz, because it corresponds to the maximum obtained for a $10\rm\, M_{\odot}$ BH; however, the emission is broadband, and further we assume that the peak is between 100 and 300 Hz for such a BH, following \cite{Gottlieb2024ApJ...972L...4G}. The different BH masses roughly rescale the frequency so that $f_{max}\propto R_{disc}^{-1} \propto M_{BH}^{-1}$; thus, we assume this relation to compute the source-frame frequency band of maximum emission for other BH masses, and we rescale it with redshift of a given source to the observer-frame. As $M_{disc}$ and $R_{disc}$ should be approximately proportional to each other, the observed strain should depend primarily on the ratio of accretion disc mass and the distance.

A very turbulent accretion disc during the hyper-accretion phase should effectively produce broadband GWs. However, once the disc enters the MAD state, the accretion disc is expected to be smoother. In principle, GWs would still be produced, because the disc cannot be perfectly turbulence-free; however, their amplitude should be much smaller than in the initial state. 
We expect almost constant emission of GWs as long as the disc is in a hyper-accretion state, therefore over a duration of the prompt duration in {\bf Model A} (see second panel of Fig. \ref{fig:overview}) and before or only at the beginning of the prompt in {\bf Model B} (see third panel of Fig. \ref{fig:overview}). 
In the MAD-powered prompt phase, it is expected for the BH to produce GWs inefficiently; thus we call this behaviour the GW silent spin-down.

\subsubsection{Descending chirp}

\citet{vanPRL01} propose that the SA causes a relatively efficient conversion of the BH's rotational energy to the coherent GW emission. 
They predict that GW luminosity should mimic changes in the BH's energy; therefore, at a given stage it should drop exponentially. 
As the Innermost Stable Circular Orbit (ISCO) expands during spin-down, the torus emits GWs at frequency
decreasing with time. 

\citet{van23} report evidence for a post-merger descending-chirp signal associated with GW170817–GRB170817A, at an observational significance of $5.5\sigma$. 
Based on GW calorimetry, this signal, GW170817B, is interpreted with the spin-down of a Kerr BH. The uniqueness of the Kerr solution provides a template for similar emissions from BH-powered core-collapse supernovae \citep{vanPutten2024}.

\section{Detectability of GWs from collapsar-origin BHs} \label{sec:detectability}

Within {\bf Model A}, the authors of \cite{Gottlieb2024ApJ...972L...4G} claim that collapsar GRBs could produce GWs detectable even by the Advanced LIGO with a rate of up to 1 event per year. However, their result is based on an assumption that a relatively massive ($10\rm\, M_{\odot}$) BH would form at a small distance ($10\rm\, Mpc$). 
The model of \cite{Lenart2025} for GRB afterglow LCs provides an estimate of the BH mass right at the beginning of the plateau, and the integrated accretion rate provides an estimate of the accretion disc mass. Therefore, it is possible to check if any of the historic GRBs would have an associated observation of GW with current or planned/future instruments. 
Given that the mass of the accretion disc during the prompt phase of a GRB should be larger than at the beginning of the plateau, because some of the matter is consumed by the BH during the prompt phase, our estimates will place a lower limit on the GW strain.

\begin{figure}
    \centering
    \includegraphics[width=0.95\linewidth]{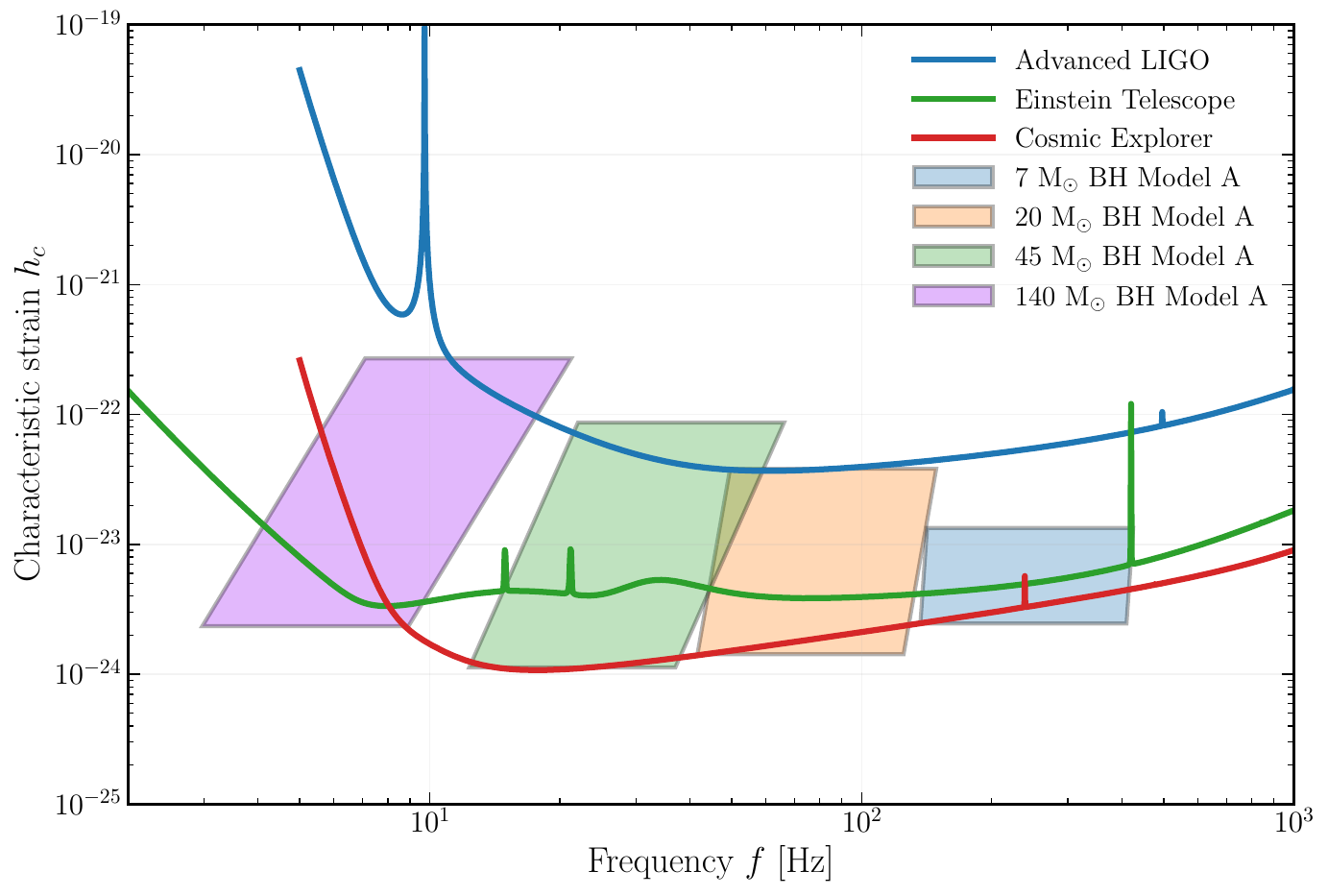}
    \caption{Sensitivity functions (unitless characteristic strain $h_c$ vs observer-frame frequency $f$ [Hz]) of Advanced LIGO, Einstein Telescope, and Cosmic Explorer, shown as blue, green, and red curves, respectively. The expected regions of GW signal from collapsar GRBs that produce BH masses of 7, 20, 45 and 140 $M_{\odot}$ are shown with blue, orange, and green parallelograms, respectively. The expected strain was calculated with the \cite{Gottlieb2024ApJ...972L...4G} model (see Sec. \ref{sec:detectability}), assuming possible redshifts between 0.0085 (the closest GRB ever detected; \cite{Galama1998Natur.395..670G}) and 0.045, 0.17, 0.65, and 1.4, which are the horizons of the strain detection by the Cosmic Explorer for BHs of masses 7, 20, 45, and 140 $M_{\odot}$, respectively.}
    \label{fig:BHvsmagnetar}
\end{figure}

We show examples of the expected characteristic strain within {\bf Model A} in Fig. \ref{fig:BHvsmagnetar}. We assumed the BH masses of 7, 20, 45, and 140 $\rm M_{\odot}$, and that the accretion disc mass is $\sim 10\%$ of the BH mass. We show the expected range of characteristic strain by assuming a range of redshifts between 0.0085 (the closest GRB ever detected; \cite{Galama1998Natur.395..670G}), and the maximum value at which the strain is comparable with the sensitivity of Cosmic Explorer. Theoretical models of stellar evolution do not predict formation of BHs with masses $\sim 140\rm M_{\odot}$ in the local Universe, because of the high metallicity of contemporary stars; however, low-metallicity galaxies do exist at low redshift \citep{Laseter2022ApJ...935..150L}. Since this is the highest mass of BH measured by LIGO \citep{Abac2025ApJ...993L..25A}, we show it for completeness.

The mass of the accretion disc is approximately proportional to the product of the accretion rate and the accretion timescale, which can be shown to be proportional to the total energy emitted during the spin-down phase. Therefore, one can search for candidates of observable GW emission by searching for the optimal ratio of $\frac{L\times T}{d}$. $L\times T$ is the plateau energy computed as plateau duration ($T$) multiplied by its luminosity ($L$), and $d$ is the proper distance. We began to calculate the strain ($h$) for the sources with the highest such quantity, until we arrived at $h_{min}=10^{-24}$, which is below the limit for Cosmic Explorer at 200 Hz. We ended up with 4 sources described in Sec. \ref{sec:data}.

The substitution of BH and disc masses estimated in Sec. \ref{sec:electromagnetic} into Eq. \ref{eq:GWs}, while assuming pessimistic $\epsilon=0.1$ resulted in characteristic strain of $(1.65\pm0.34)\times 10^{-24}$, $(1.84\pm0.74)\times 10^{-24}$, $(1.5\pm0.74)\times 10^{-24}$, for GRBs 080319C, 111209A and 130925A, respectively (see Tab. \ref{tab:fitting}). The uncertainty was calculated with the statistical uncertainties on BH parameter estimates, and we additionally assumed 50\% of the systematic uncertainty. The statistical uncertainty is relatively small; thus, we added this arbitrary systematic uncertainty to show on the plot the possible range of expected characteristic strain.

We present these estimates compared with the detector sensitivity in Fig. \ref{fig:sensitivity}. We estimated the sensitivity curves using the \cite{Robson2019} code for LISA (Laser Interferometer Space Antenna; \cite{Colpi2024arXiv240207571C}), the ASD sensitivity curves repository \citep{LIGO-T1500293} for Advanced LIGO (Laser Interferometer Gravitational-Wave Observatory; \cite{Abbott2016PhRvL.116m1103A}), Einstein Telescope \citep{Abac2026JCAP...03..081A}, and Cosmic Explorer \citep{Reitze2019BAAS...51g..35R}, and the \cite{Yagi2011PhysRevD.83.044011} formulas for DECIGO (DECi-hertz Interferometer Gravitational wave Observatory) and BBO (Big Bang Observer).

Therefore, in this pessimistic scenario, GRBs 111209A and 130925A would be on the verge of detection by the Cosmic Explorer. 

However, given that our accretion disc estimates are possibly underestimated, and higher GW efficiency production is possible, all three sources could be detectable by the Cosmic Explorer. Assuming the most optimistic scenario $\epsilon\approx0.3$ \citep{Gottlieb2024ApJ...972L...4G} and the disk mass being $~40\%$ of the initial BH mass \citep{ShibataPhysRevD.109.043051,Shibata2026ApJ...996...57S}, the resulting characteristic strain is $(1.98\pm0.41)\times 10^{-23}$, $(2.21\pm0.89)\times 10^{-23}$, $(1.8\pm0.9)\times 10^{-23}$, for GRBs 080319C, 111209A and 130925A, respectively (see dashed lines in Fig. \ref{fig:sensitivity}).

Given that Swift's BAT field of view is $\sim1.4$ sr, and our 3-source sample was gathered over $\sim 18$ years, out of which two should be well detected by the Cosmic Explorer, and one just outside its frequency band, we estimate that this instrument could be able to detect $\sim 1$ LGRB plateau-associated broadband emission annually (in the optimistic scenario $\epsilon=0.2$).

In the highly optimistic scenario, we would need to analyse a higher number of GRBs to give a precise estimate of the observed rate. However, both Cosmic Explorer and Einstein Telescope should be able to detect a few ($>1.5$) such events annually.
 
In the pessimistic scenario, the rate would be $\sim 0.5$ such events per year ($\epsilon=0.1$). 

Taking into consideration that a large portion of GRBs might produce jets not towards the Earth, there might be a large number of GW events without accompanying GRB observation. Also, GRBs without a plateau can be associated with the creation of a BH undergoing hyperaccretion; thus, such events may also provide a higher detection rate.

One needs to remember that the broadband emission might be hard to detect and measure, because of its noise-like nature \citep{AghaeiAbchouyeh_2025}.

Although we do not make specific calculations in this paper, within the GW-loud BH spin-down in {\bf Model B} of \cite{vanPutten2001b,vanPutten2003,vanPutten2004}, GWs should be more readily detectable.
They consume a few \% of the BH's rotational energy \citep{vanPutten2024}, thus should be much brighter than a broadband emission from the disc, 
and in such a scenario we expect a descending chirp-like emission \citep{AghaeiAbchouyeh2026ApJ..1006..212A}.

\section{Constraints on the physics of the LGRBs}
\label{sec:constraints}

\subsection{TDE vs Collapsars in relation to ULGRBs}
\label{sec:tde}


Recently, it was proposed that ULGRBs are powered by TDEs \citep{Krolik2011ApJ...743..134K,MacLeod_2014,Levan2014,Levan2015JHEAp...7...44L,Ioka2016ApJ...833..110I}. As we discuss in this section, such a scenario would lead to the production of GWs possibly detectable by space-based detectors in the band $ < 0.1$ Hz; therefore, at frequencies different from the ones expected from a collapsar GRB (see discussion later in the text). The detection of such a signal would confirm the TDE origin of at least some ULGRBs.

\begin{figure}
    \centering
    \includegraphics[width=0.95\linewidth]{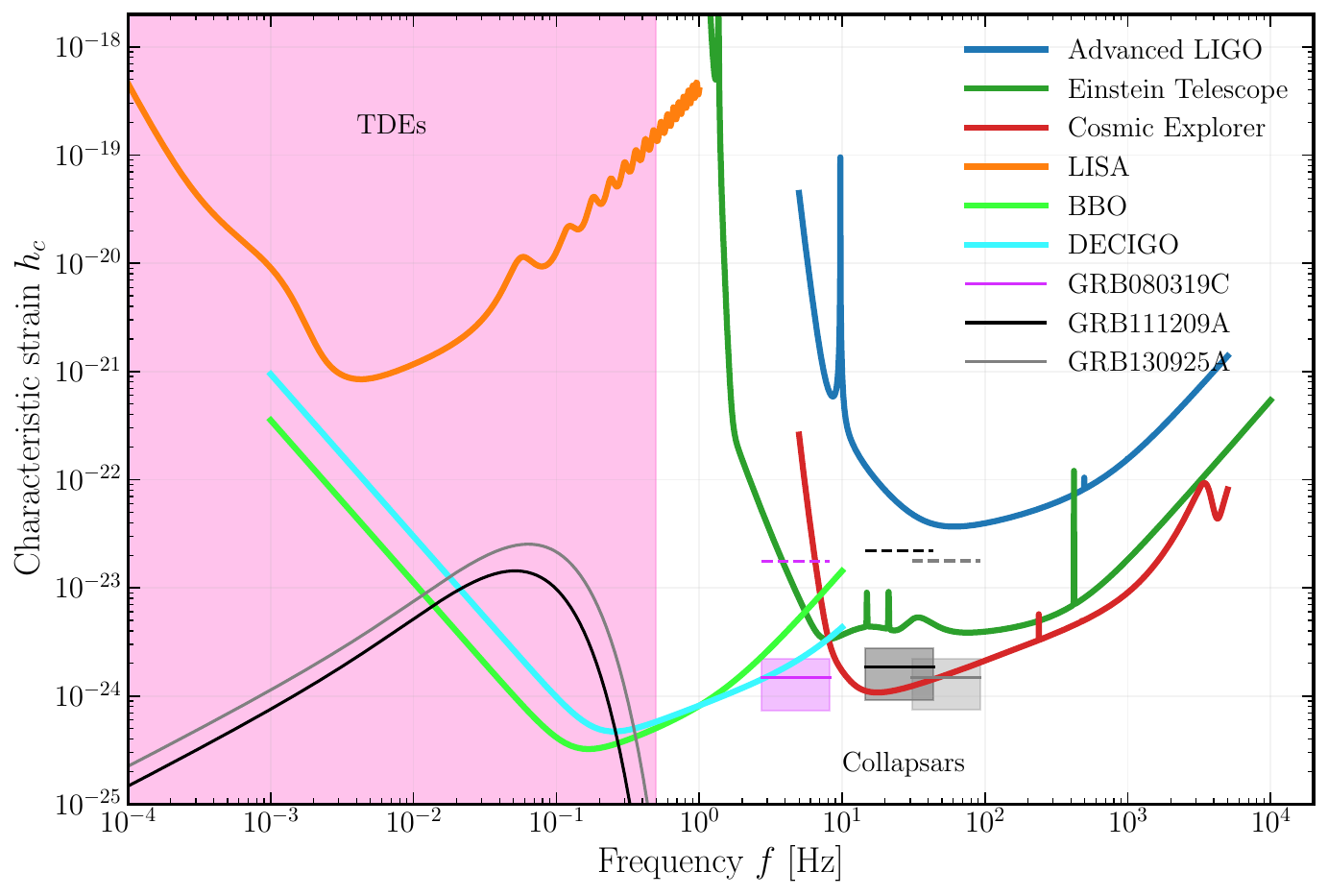}
    \caption{Sensitivity functions (unitless characteristic strain $h_c$ vs observer-frame frequency $f$ [Hz]) of Advanced LIGO, Einstein Telescope, Cosmic Explorer, LISA, BBO and DECIGO, shown as blue, dark green, red, orange, light green and cyan curves, respectively. The grey, black, and magenta lines show the expected GW signals from the GRBs in our sample.
    For the frequencies $<$0.1 Hz (pink region), we show the expected GW emission from TDEs. Dashed lines correspond to the parabolic plunge scenario, while continuous lines correspond to the scenario of multiple signal accumulation over multiple orbital periods (see Sec. \ref{sec:tde}). For the frequencies $>$0.1 Hz, we show the expected strain emitted by GRBs in our sample in the case of the \cite{Gottlieb2024ApJ...972L...4G} model (see Sec. \ref{sec:detectability}). Continuous lines correspond to pessimistic estimates discussed in Sec. \ref{sec:detectability}, while dashed lines above them correspond to a more optimistic scenario. Shaded rectangles correspond to 1$\sigma$ confidence intervals including a conservative estimate of 50\% systematic uncertainties; see Sec. \ref{sec:detectability}.}
    \label{fig:sensitivity}
\end{figure}

We found that a TDE involving a white dwarf and a $10^5\,\rm M_{\odot}$ BH at the distance of ULGRBs 130925A and 111209A could be detected by the DECIGO and BBO GW observatories; see Sec. \ref{sec:TDE} for detailed calculations and Fig. \ref{fig:sensitivity}.

We note here that different ULGRBs from those in our sample, 060218 and 100316D, were observed at redshifts 0.0331 and 0.0591, respectively. 
Thus, if produced in similar TDEs, their strain should be roughly an order of magnitude higher than our prediction. If the WD would complete many orbital cycles close to the ISCO before being disrupted, such events could possibly be on the verge of LISA's sensitivity.

\subsection{Magnetar vs BH central engine}

GRB magnetar models usually assume that the accretion cannot be too high, because when the NS reaches a critical mass it becomes unstable and collapses to a BH. This would likely cause a sudden drop in the electromagnetic emission. A similar model was proposed to explain a few SGRB LCs \citep{Lu2015}. However, this should not happen for a typical afterglow. Therefore, we do not expect efficient emission of GWs from the accretion disc surrounding the magnetar. In the magnetar scenario, the GW emission would likely come from bar-mode oscillations, which produce a quasi-monochromatic spectrum in the frequency range of LIGO ({\bf Model C} in Fig. \ref{fig:overview}). 
Initially, during the GRB prompt phase, the magnetar is spinning up and gains mass; thus, its ellipticity can grow, and the magnetar can produce stronger GWs. The maximum of the GW emission should be reached right after the prompt, when the accretion becomes inefficient and energy-extraction processes dominate, such as dipole electromagnetic radiation. 
Then the magnetar is spinning down, and the characteristic strain of the GWs becomes smaller. Thus, contrary to BHs, where the peak of GW emission should be during or before the prompt phase, magnetars will produce the peak right after the prompt. Given the smaller mass of NS, the expected strain should also be much smaller than in the case of BH.
The magnetar also should produce a descending chirp-like emission, given that the frequency of GWs is proportional to the frequency of NSs rotation, which is slowing down over time. The rotation period of the neutron star measured with the magnetar model of the afterglow \citep{Rowlinson2014}, distance, the strain, and timing will be discriminating for the model, given there is a known limit on maximum neutron star mass \citep{dallosso2018}.

\subsection{Type of Accretion on the Collapsar-origin BH}

The timing and morphology of GW emission could help distinguish between Models A and B (Sec. \ref{sec:model}) and identify which processes occur at each stage of a GRB. 
Within {\bf Model A} we expect a broadband emission over the whole prompt duration, while in {\bf Model B} we expect it only right before the prompt. In the latter model, it is also possible to detect a descending chirp-like emission over the prompt duration. The summary is presented in Fig. \ref{fig:overview}.

\section{Summary and Conclusions}
\label{sec:conclusions}

We have examined what GWs can reveal about the central engines of long GRBs with X-ray plateaus, combining a physical model of the electromagnetic afterglow with GW emission scenarios for collapsar BHs, magnetars, and TDEs. Our main results are as follows: 

\begin{enumerate}
\item \textit{Three engine scenarios make distinct GW predictions.} We distinguish ({\bf A}) a prompt phase powered by hyper-accretion followed by a plateau powered by BH spin-down in the MAD state; ({\bf B}) a prompt phase powered
by BH spin-down, in which the plateau comes from the residual disc or a less luminous jet; and ({\bf C}) a magnetar central engine. In {\bf Model A}, broadband, stochastic-like emission from the turbulent disc \citep{Gottlieb2024ApJ...972L...4G} is expected throughout the prompt phase. In {\bf Model B}, such emission is confined to a short hyper-accretion phase before the prompt, and in the SA case the spin-down itself can produce a coherent descending chirp. In {\bf Model C}, the signal is quasi-monochromatic, peaks shortly after the prompt phase, and declines as the neutron star spins down. The timing of the GW emission relative to the prompt and plateau, together with its spectral morphology, is therefore the key discriminant between these scenarios (Fig.~\ref{fig:overview}).

\item \textit{Plateau analysis within the {\bf Model A} provides BH and disc masses.} Fitting the BH spin-down model of \citet{Lenart2025} to the X-ray LCs of the three GRBs in our sample for which the fit converged gives initial BH masses of $\sim$24--124\,$M_\odot$ and disc masses of $\sim$3--12\,$M_\odot$ at the start of the plateau (Tab. ~\ref{tab:fitting}). Because the disc mass does not suddenly gain mass after the transition to the MAD stage, these values describe the minimum disc at the end of the prompt phase and are lower limits on its mass earlier in the prompt.

\item \textit{The brightest events lie near the sensitivity of third-generation detectors.} Evaluating the \citet{Gottlieb2024ApJ...972L...4G} scaling for our BH and disc masses with efficiency $\epsilon = 0.3$ and highly massive discs ($0.4\,M_{BH}$) gives $h_{c} \approx 1.98\times10^{-23}$ for GRB~080319C and $\approx 2.21\times10^{-23}$ for GRBs~111209A and $\approx 1.8\times10^{-23}$ for 130925A. While with the pessimistic efficiency $\epsilon = 0.1$ gives $h_{200\,\mathrm{Hz}} \approx 1.7\times10^{-24}$ for GRB~080319C and $\approx 1.8\times10^{-24}$ for GRBs~111209A and $\approx 1.5\times10^{-24}$ for 130925A. These values are far below the reach of Advanced LIGO, but are comparable to the Cosmic Explorer sensitivity near 200\, Hz, so GRBs~111209A and 130925A are on the verge of detectability in this pessimistic case, and all three become accessible for a higher efficiency and heavier disc. Assuming that our sample is representative, we estimate a rate of $\sim$0.5 -- $>$1.5 detectable events per year for Cosmic Explorer, with additional GRB-less events expected in case of off-axis jets. These estimates carry large uncertainties, since the sample is small and the signal is noise-like and therefore harder to extract than a matched-filter signal.

\item \textit{GWs can test the origin of ULGRBs.} Of our sample, three events are commonly classified as ULGRBs. If such events are powered by TDEs of white dwarfs by intermediate-mass BHs rather than by collapsars, the expected GWs fall at mHz--dHz frequencies, well below the expected collapsar band. A single passage lies below the LISA sensitivity; however, it should be detectable by DECIGO and BBO, whereas a bound orbit accumulating many cycles could be seen by LISA. The two bands thus provide an unambiguous distinction between the scenarios.

\item \textit{GWs can separate BH and magnetar engines.} BH engines produce their GW peak during the prompt phase, whereas magnetars peak shortly afterwards, with a much lower amplitude given the smaller mass of the neutron star. Both can show a descending chirp, but with different shapes and timescales, so the morphology and the timing of the chirp can place strong limits.
\end{enumerate}

This work shows how crucial future GW detectors are for the development of the fields of GRBs and BHs. Detectors like Cosmic Explorer, Einstein Telescope, DECIGO, and BBO could provide us with paradigm-changing observations. Even upper limits could help to resolve long-standing degeneracies between GRB models.

Our results also imply that the third-generation GW detectors will open a new chapter in the accretion disc physics, as they will discriminate some models.

Our estimates of GW detectability are primarily based on the numerical results of \cite{Gottlieb2024ApJ...972L...4G} and a comparison of the predicted characteristic strains with detector sensitivity curves. A more quantitative assessment, including signal-to-noise ratios, detector responses, and search efficiencies, is beyond the scope of this study and represents an important direction for future work.
Our results rest on a small sample, as this is a case study; however, the growing number of GRB observations with well-sampled plateaus and consideration of more optimistic scenarios might enable the prediction of higher numbers of GRBs detectable together with GWs in the future.

Third-generation detectors may have the potential to detect broadband disk emission with the presented further
identification of hyper-accreting or MAD states \citet{vanPutten2024,AghaeiAbchouyeh2026ApJ..1006..212A}, 
magnetar central engines and TDEs.

\begin{acknowledgments}

The study was funded by the Research Support Module as part of the "Excellence Initiative - Research University" program at the Jagiellonian University in Kraków.
M.G.D. and S.N. acknowledge support from the JSPS Grant-in-Aid for Scientific Research (KAKENHI) (A), Grant Number JP25H00675. S.N. also acknowledges support from JSPS KAKENHI (B), Grant Number JP23K25874, and the JST ASPIRE Program, 'RIKEN–Berkeley Mathematical Quantum Science Initiative. A.Ł.L. and M.v.P. acknowledge the support of NAOJ for the research stay at Division of Science where the idea was developed and the paper was written. 
M.v.P. acknowledges support from the Korean NRF RS-2024-00334550. N.F.  acknowledges  financial  support  from UNAM-DGAPA-PAPIIT  through  grant  IN112525.
We thank Prof. Daniel Kasen for fruitful discussions about the magnetar model and the several configurations of BH masses adopted in Fig. \ref{fig:BHvsmagnetar}. 

\end{acknowledgments}

\begin{contribution}

Idea development (AŁL, MGD, MvP, DHH, SN), text writing (AŁL, MvP, MGD), preparation of figures and calculations (AŁL), text revision and discussion (all authors).

The authors declare the use of generative AI in the research and writing process. According to the GAIDeT taxonomy (2026), the following tasks were partially delegated to GAI tools under full human supervision: literature search and systematization, evaluation of the novelty of the research and identification of gaps, proofreading and editing, reformatting, quality assessment, identification of limitations.

The GAI tool used was Claude Sonnet 5.5.
Responsibility for the final manuscript lies entirely with the authors.
GAI tools are not listed as authors and do not bear responsibility for the final outcomes.
Declaration submitted by: Collective responsibility
Additional note: We used Claude to enhance the visual side of the plots.


\end{contribution}

%
\facilities{Swift(XRT)}

\software{astropy \citep{2013A&A...558A..33A,2018AJ....156..123A,2022ApJ...935..167A}
          }


\appendix

\section{Electromagnetic estimates on properties of BHs formed in collapsars} \label{sec:electromagnetic}

In order to discuss the detectability of GWs from GRBs, one needs to understand their possible properties. We discuss electromagnetic estimates in this section.

\subsection{Constraints from the GRB data}

\begin{figure}
    \centering
    \includegraphics[width=0.32\textwidth]{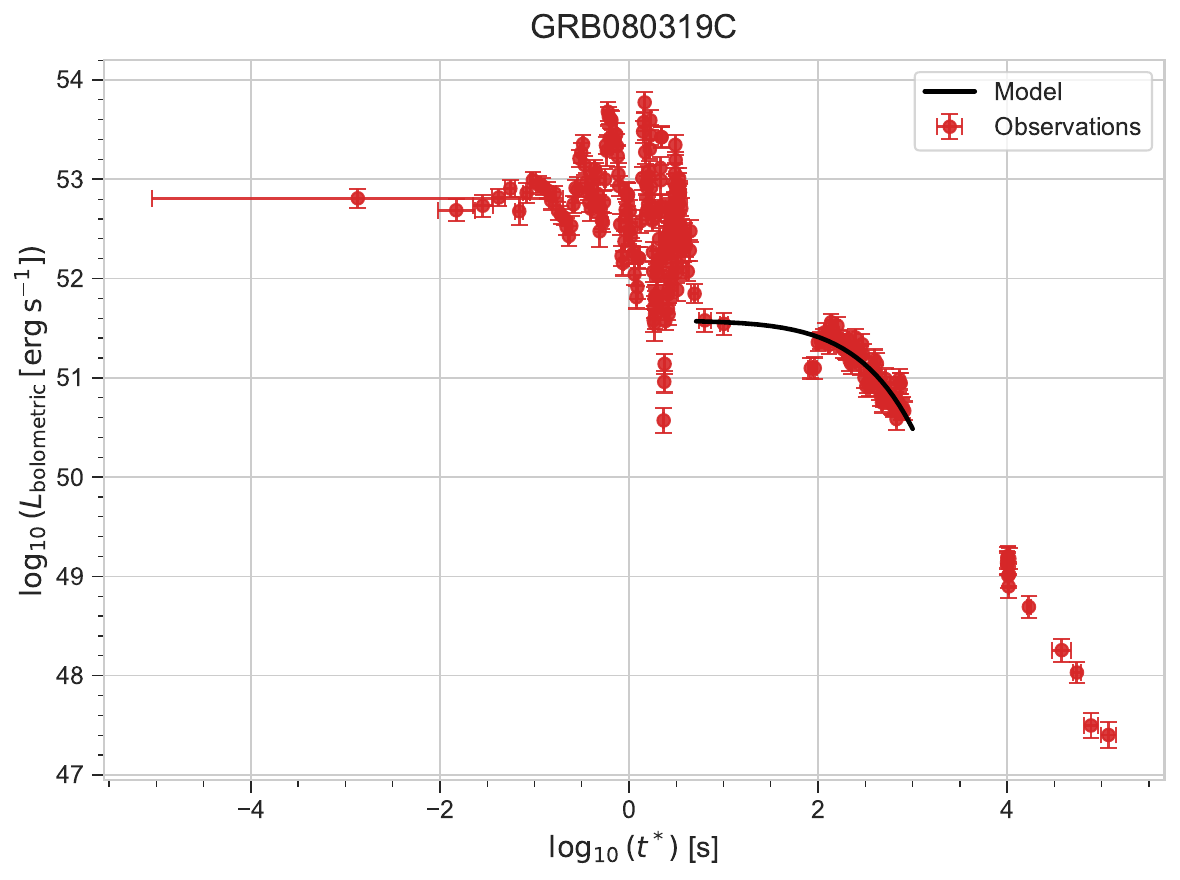}
    \includegraphics[width=0.32\textwidth]{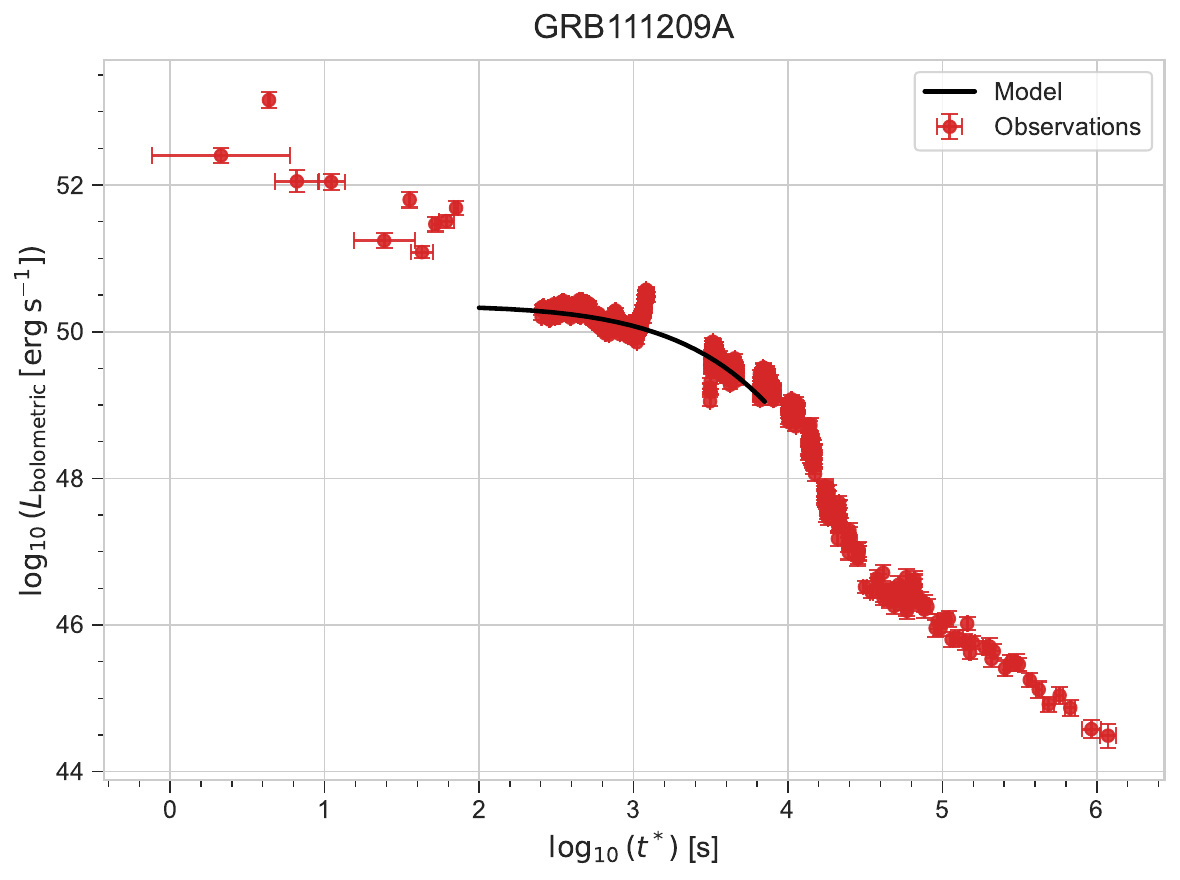}
    \includegraphics[width=0.32\textwidth]{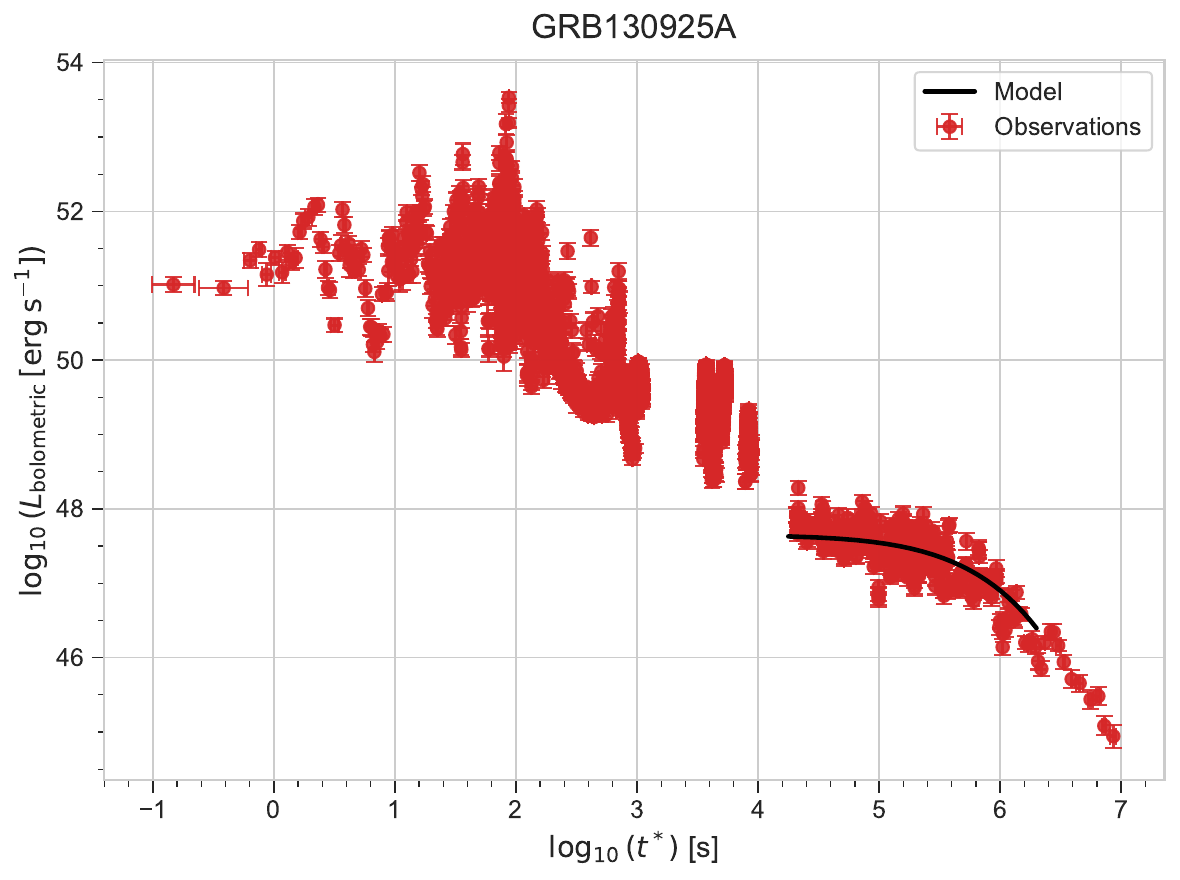}
    \includegraphics[width=0.32\textwidth]{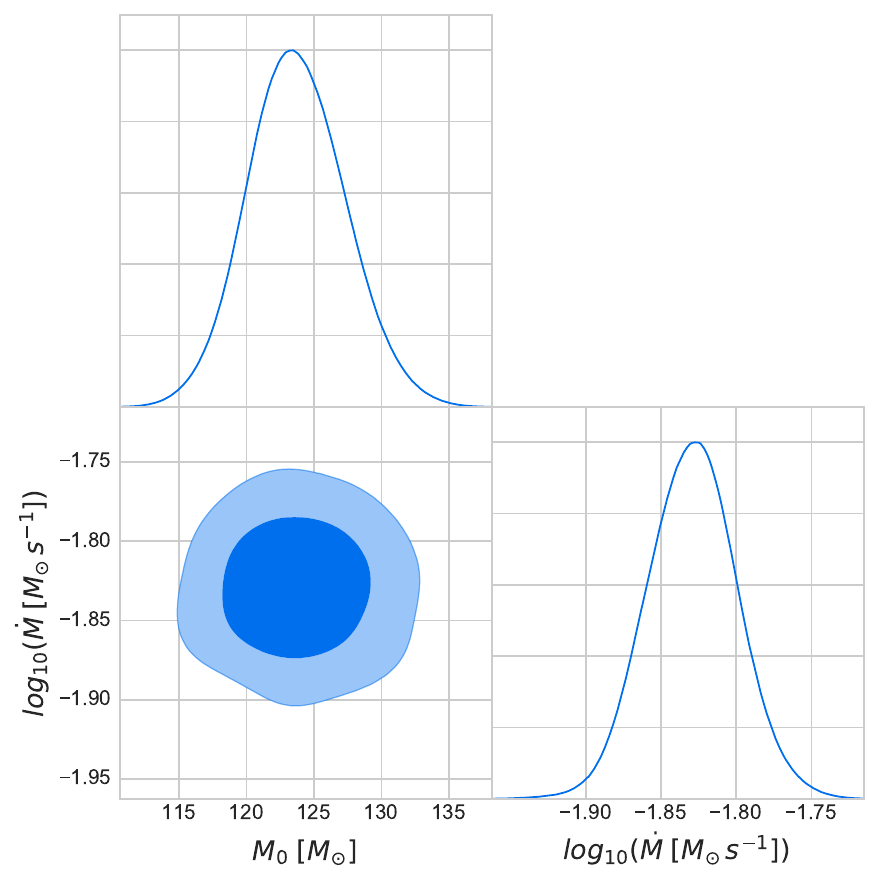}
    \includegraphics[width=0.32\textwidth]{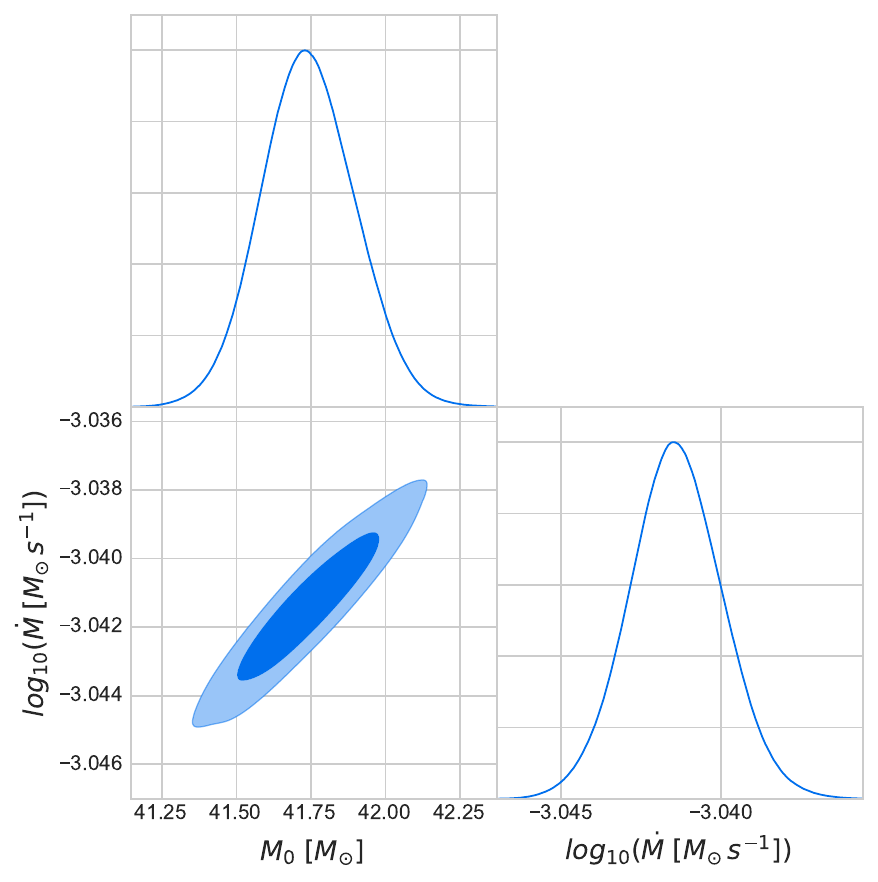}
    \caption{Upper panels: the LC data (red points) and the best-fit model function (black curves) of GRBs 080319C, 111209A, and 130925A, respectively. The model is plotted only over the period of the fitting reported in Tab. \ref{tab:fitting}. Lower panels: the MCMC posterior parameter space for GRBs 080319C and 111209A, respectively.}
    \label{fig:grblc}
\end{figure}

We performed an MCMC estimation of the parameters of the BH from GRB afterglow LC within {\bf Model A} (described in more detail by \cite{Lenart2025}) using the data of GRBs from our sample. We assumed that the X-ray luminosity is $\sim10\%$ of the bolometric luminosity, and that the beaming and radiative efficiency cancel out. 
Given that this model neglects the shock contribution, we can only apply it to the data up to right after the break in the LC. We report the periods used in the fitting in Tab. \ref{tab:fitting}. The posteriors of the fitting are shown in Fig. \ref{fig:grblc}, and the mean values of parameters from chains are shown in Tab. \ref{tab:fitting}. 
The quoted uncertainties represent statistical uncertainties obtained from the MCMC analysis within the adopted model assumptions calculated as the standard deviations of the MCMC posteriors. They do not include systematic uncertainties associated with the assumed radiative efficiency, beaming factor, bolometric correction, or accretion history. The inferred BH masses should therefore be regarded as model-dependent estimates.
The fitted functions and the corresponding data are shown in Fig. \ref{fig:grblc}. For GRB 130925A, we fitted the model to the data. However, MCMC failed due to very high scatter; thus, we do not report uncertainties on the BH parameters, but for completeness we add this GRB to our sample.
For GRB 211024B, the fitting failed as it was impossible for the algorithm to find the likelihood maximum. 

\begin{table}[ht!]
    \centering
    \begin{tabular}{|c|c|c|c|c|c|c|c|}
    \hline
        GRB & $\rm log_{10}t_{start}$ -- $log_{10}t_{end}$ & $M_0\,\rm [M_{\odot}]$ & $\log_{10}(\dot{M}\,\rm [M_{\odot}\, s^{-1}])$ & $M_{disc}\,\rm [M_{\odot}]$ & $z$ & $h_{200\,\rm Hz}$ \\\hline\hline
        080319C & [0.71, 3] & $123.5\pm 3.7$ & $-1.83\pm 0.03$ & $12.0\pm 0.8$ & 1.95 & $(1.65\pm0.34)\times 10^{-24}$ \\\hline
        111209A & $[2,\, 3.02] \cup [3.7,\,3.85]$ & $41.74\pm 0.16$ & $-3.041\pm 0.002$ & $6.43\pm 0.02$ & 0.677 & $(1.84\pm 0.74)\times 10^{-24}$ \\\hline
        130925A & [4.25-6.3] & $24.0$ & $-5.8$ & $2.9$ & 0.347 & $(1.5\pm0.74)\times 10^{-24}$ \\\hline
    \end{tabular}
    \caption{For each GRB listed in the first column, the second column gives the time intervals used in the fit.
    The best-fit model parameters, initial BH mass $M_0$ in units of $\rm M_{\odot}$, and the log accretion rate $\dot{M}$ in units of $\rm M_{\odot}s^{-1}$ are in the third and fourth columns, respectively. The disc mass estimated as an integrated $\dot{M}$ is shown in the fifth column, which is followed by the observed redshift $z$ in the sixth column and the characteristic strain estimated with Eq. \ref{eq:GWs} in the seventh column.}
    \label{tab:fitting}
\end{table}

We stress here that our model predicts only the BH mass at the beginning of the plateau (end of prompt), and assumes the accretion rate to be constant before and during the MAD stage. Thus, the BH mass does not match the mass right after the core collapse. The BH could have grown significantly during the prompt due to the hyper-accretion. The disc mass is estimated as the accretion rate integrated over the time from $t=0$ to the time of the end of the fitting period. Thus, the disc mass might be underestimated. Moreover, it is important to note that the disc mass at the beginning of the prompt emission should be significantly larger if the prompt emission is powered by accretion.

The biggest obtained BH mass, $123.5\,M_{\odot}$, may seem high, but is theoretically possible for a GRB. Several studies have suggested that exceptionally massive GRB progenitors, particularly rapidly rotating Population III stars, may collapse to form BHs with masses exceeding $100\,\rm M_\odot$ and potentially reaching the intermediate-mass regime \citep{Komissarov2010,Suwa2011,Gottlieb2025ApJ...993L..54G,Renzo2026arXiv260621824R}. \cite{Guglielmo2024ApJ...961...89V} discusses the production of a BH of mass reaching even $\sim 170\,\rm M_{\odot}$ from a very small metallicity $200\,\rm M_{\odot}$ star. In the ``supercollapsar'' scenario, metal-free stars with initial masses of $\gtrsim 10^2$--$10^3\rm\,M_\odot$ can power long-duration relativistic jets while leaving behind BHs massive enough to act as seeds of supermassive BHs \citep{Komissarov2010}. Such a scenario has been proposed as a potential origin of ULGRBs, with prompt emission lasting \(10^4\)–\(10^5\) s due to prolonged accretion onto a newly formed BH of \(\gtrsim 100\,M_\odot\) \citep{Komissarov2010,Suwa2011,Nakauchi2012}. In fact, 3 out of 4 GRBs (except 080319C) in our sample are often classified in the literature as ULGRBs, eg., \citep{Levan2015JHEAp...7...44L,Fu2024ApJ...977..197F}. In our Tab. \ref{tab:fitting} we report smaller values of $T_{90}$ than the usual definition of the ULGRB class; however, the authors of the aforementioned articles argue that values measured by {\it Swift} BAT are not representative and give their own estimates of $T_{90}\sim 2.5\times 10^4$ s for 111209A \citep{Gendre2013}, $T_{90}\sim 2\times 10^4$ s for 130925A \citep{Schady2015A&A...579A.126S} and $T_{90}\sim 1.5\times 10^4$ s for 211024B \citep{Fu2024ApJ...977..197F}.
However, in the case of GRB 111309A, the $T_{90}$ estimate is based on the interpretation of the X-ray component as part of the prompt emission, although it does not show typical prompt variability. The X-ray component fitted by us was classified as the plateau phase by \cite{Narendra2024}, which builds on previous samples \cite{Dainotti2023c,Dainotti2021ApJS..255...13D,Dainotti2020ApJ...904...97D,dainotti2017}  based on the fitting by the \cite{Willingale2007} model.

We note that the late-time X-ray emission of GRB 130925A has also been interpreted as a dust-scattering echo of the prompt emission \citep{Evans2014_10.1093/mnras/stu1459}. In the present study, we explore the alternative hypothesis that the X-ray component considered in our analysis is predominantly powered by sustained central-engine activity. The inferred black-hole and accretion-disk parameters should therefore be interpreted within this assumption.

\subsection{Constraints from the optical data}

The possible SN association of a GRB sets a constraint on the produced BH mass \citep{AghaeiAbchouyeh2026ApJ..1006..212A}. Type Ic supernovae, which come from stars stripped of their hydrogen and helium envelopes, have inferred ejecta masses of only $\sim$1--10 $M_\odot$ \citep{Drout2011, Lyman2016,Cano2013,Taddia2019}, which implies CO-core progenitors of roughly 10--20 $M_\odot$ and compact remnants of at most a few to $\sim$15 $M_\odot$ \citep{Sukhbold2016ApJ...821...38S}. Hydrogen-poor superluminous SNe, powered by pair-instability pulsations and ejecta interaction, probe much more massive progenitors, with ejecta masses reaching $\sim$0.13--30 $M_\odot$ and He/CO cores in the pulsational pair-instability range, so any BHs they leave behind are expected to be $\lesssim$30--50 $M_\odot$ \citep{Woosley2017}. 

We also note the existence of supernova-less LGRBs such as GRB 060505 and GRB 060614, which show no associated SN down to limits hundreds of times fainter than SN 1998bw, which points either to a failed explosion with direct black-hole formation and little ${}^{56}$Ni (the original collapsar picture) produced. Therefore, such events may form higher-mass BHs. It should also be noted that such GRBs may be of a compact-binary merger origin \citep{Fynbo2006, GalYam2006, DellaValle2006, Gehrels2006, Ofek2007ApJ...662.1129O}.

No SN was observed associated with GRB 080319C; however, given its high redshift $z = 1.95$, this is not constraining, since SN would likely be too faint.
The SN\,2011kl event associated with GRB\,111209A represents one of the most luminous SNe ever discovered in association with a GRB. Spectroscopic and photometric analyses showed that SN\,2011kl was substantially more luminous and significantly bluer than typical broad-lined Type Ic SNe associated with LGRBs, while at the same time remaining less luminous than classical superluminous SNe \citep{Greiner2015,Kann2019}. Modelling of the bolometric LC indicated that reproducing the observed luminosity through radioactive heating alone would require an unusually large amount of $^{56}$Ni ($2.27\pm0.64\,\rm M_{\odot}$), in tension with the observed spectra \citep{Greiner2015,Kann2019}. Consequently, these works favored an additional central-engine power source. The inferred ejecta properties place SN\,2011kl among the most energetic GRB-SNe. The modelling by \citet{Greiner2015} found an ejecta mass of $M_{ej} = 6.79^{+3.67}_{-2.84}$ solar masses together with a kinetic energy of order $10^{52}$ erg. 
Subsequent analysis by \citet{Kann2019} confirmed the exceptional luminosity and spectral properties of SN\,2011kl, arguing that it bridges the gap between ordinary GRB-SNe and superluminous supernovae. These results imply the collapse of a massive stripped stellar core and the formation of a highly energetic compact remnant. While the supernova observations alone do not uniquely distinguish between a magnetar and BH central engine, the large explosion energy and ejecta mass are consistent with progenitor masses capable of producing a several-tens-of-solar-mass BH in a pulsational pair-instability SN \citep{Woosley2017}.

It was argued that the presence of this SN excludes the TDE origin of this source \citep{Greiner2015}. However, \cite{Ioka2016ApJ...833..110I} argues that a TDE model can explain its peculiar nature.
In addition, this source was found close to the nucleus of its galaxy, which supports the TDE model \citep{Levan2014}.

GRB\,130925A exhibited highly unusual optical and near-infrared behaviour. Detailed GROND observations revealed a bright optical/NIR flare delayed by approximately 300--400\,s relative to the high-energy emission, with an amplitude approaching two magnitudes and a remarkably rapid decay \citep{Greiner2014}. The observed variability could not be straightforwardly explained within standard prompt-emission or external-shock afterglow scenarios, suggesting a more complex origin related to prolonged central-engine activity or internal dissipation processes \citep{Greiner2014}. The unusual optical evolution distinguishes GRB\,130925A from more typical LGRBs and supports the view that ULGRBs may arise from progenitors or engine configurations different from those operating in ordinary collapsars. Unlike GRB\,111209A/SN\,2011kl, however, no well-characterized SN counterpart has been found. This source was sufficiently close for a typical LGRB-associated SN to be detected; however, high extinction was detected, which might be the reason for the lack of observation \citep{Greiner2014}. Consequently, GRB\,130925A provides only limited constraints on the mass of the compact remnant through supernova observations alone \citep{Greiner2014,Evans2014}.

\cite{Schady2015A&A...579A.126S} measure the metallicity of this burst's host, obtaining a super-solar value. This source was also localized close to its host nucleus. Those two factors seem to support the TDE origin.

\section{Tidal Disruption Events}
\label{sec:TDE}

We assume that a GRB event caused by a TDE involved a white dwarf with mass $m_{WD}=0.6\,M_\odot$ (radius $R_{\rm WD}\simeq 8.7\times10^{8}$\,cm) disrupted on a parabolic orbit with penetration factor $\beta\simeq1$ (pericenter distance comparable with the tidal radius) by a $10^{5}\,M_\odot$ BH (upper limit from \cite{Levan2014}). 
The tidal radius for a non-spinning BH is given by \cite{Hills1975,Rees1988}:

\begin{equation}
r_t = R_{\rm WD}\left(\frac{M_{\rm BH}}{m_{\rm WD}}\right)^{1/3}\simeq 3.2\,\frac{GM_{\rm BH}}{c^{2}}.
\end{equation}

Following the quadrupole treatment of GW emission from stellar disruptions \citep{Turner1977,Kobayashi2004,Rosswog2009}, the burst peaks near the inverse pericentre passage time, in the source rest frame \citep{Chen2026ApJ..1009...44C}:

\begin{equation}
f_{\rm peak}\simeq 1.6f_c = 1.6\frac{1}{2\pi}\sqrt{\frac{GM_{\rm BH}}{r_{\rm t}^{3}}},
\end{equation}

where $f_c$ is the frequency of the circular orbit at $r_t$. The observer-frame characteristic strain $h_c$ is given by:

\begin{equation}
h_c(f)=\frac{1+z}{D_L}\sqrt{\frac{2G}{\pi^{2}c^{3}}\frac{dE}{df}\bigg|_{f(1+z)}}
\end{equation}

\citep{Flanagan1998,Maggiore2000,Moore2015}, and the energy derivative for the passage close to the tidal radius is given by:

\begin{equation}
    \frac{dE}{df} \approx \frac{4\pi^2}{5}\frac{G^3m_{WD}^{2}M_{BH}^2}{c^4r_{t}^2}\, l \left(\frac{f}{f_c}\right),
\end{equation}

where $l \left(\frac{f}{f_c}\right)$ is an analytical function found by \cite{Berry2010PhysRevD.82.107501}. 
Evaluating the above equations for GRBs 130925A and 111209A yields the spectra shown in Fig. \ref{fig:sensitivity}.
Such a signal lies below the LISA sensitivity \citep{Robson2019} but should be detectable with decihertz detectors such as DECIGO or BBO \citep{Kawamura2011,Harry2006}, even if the BH were less massive than we assumed.

The strain can be substantially larger for different orbital configurations. A deeper plunge ($\beta>1$) increases the amplitude only by a factor of a few, since $h\propto \beta$ \citep{Peters1963,Turner1977}.
A much larger gain comes from a bound, quasi-circular or eccentric orbit that completes many cycles in band: the signal then accumulates roughly as $h_c\sim h\sqrt{N_{\rm cyc}}$ \citep{Finn2000,Barack2004,Moore2015}. For example, $N_{\rm cyc}\sim10^{5}$ would place the GRB 130925A close to the LISA band. However, for a non-spinning BH, the tidal radius lies inside the ISCO ($6\,GM_{\rm BH}/c^{2}$; \citealp{Bardeen1972}), so the white dwarf is swallowed whole and the event is an intermediate-mass-ratio inspiral rather than a tidal disruption. Disruption near the ISCO requires a rapidly spinning BH on a prograde orbit \citep{Ivanov2006,Kesden2012}, while repeated partial disruptions can occur on eccentric orbits with pericenter radius just above $r_t$. 












\bibliography{bibliography}{}
\bibliographystyle{aasjournalv7}



\end{document}